\documentclass[prl,aps,amssymb,twocolumn,showpacs,floatfix]{revtex4-1}
\usepackage[dvips]{graphicx}
\def\be{\begin{equation}}
\def\ee{\end{equation}}
\def\bea{\begin{eqnarray}}
\def\eea{\end{eqnarray}}

\newcommand{\req}[1]{Eq.\,(\ref{#1})}

\newcommand{\rfig}[1]{Fig.\,\ref{#1}}

\begin{document}

\title{Emergence of domains and nonlinear transport in the zero-resistance state}
\author{I. A.~Dmitriev$^{1,2,3,4}$}
\author{M. Khodas$^{5}$}
\author{A. D.~Mirlin$^{2,3,6}$}
\author{D. G.~Polyakov$^{3}$}
\affiliation{$^{1}$Max Planck Institute for Solid State Research, 70569 Stuttgart, Germany,\\
$^{2}$Institut f\"ur Theorie der kondensierten Materie and DFG Center for Functional Nanostructures,
Karlsruhe Institute of Technology, 76128 Karlsruhe, Germany\\
$^{3}$Institut f\"ur Nanotechnologie, Karlsruhe Institute of Technology, 76021 Karlsruhe, Germany\\
$^{4}$Ioffe Physical Technical Institute, 194021 St.Petersburg, Russia\\
$^{5}$Department of Physics and Astronomy, University of Iowa, Iowa City, IA 52242, USA\\
$^{6}$Petersburg Nuclear Physics Institute, 188300 St.Petersburg, Russia
}

\date{\today}

\begin{abstract}
We study transport in the domain state, the so-called zero-resistance state, that emerges in a two-dimensional electron system in which the combined action of microwave radiation and magnetic field produces a negative absolute conductivity. We show that the voltage-biased system has a rich phase diagram in the system size and voltage plane, with second- and first-order transitions between the domain and homogeneous states for small and large voltages, respectively. We find the residual negative dissipative resistance in the stable domain state.
\end{abstract}


\maketitle

\noindent
{\it Introduction.---}The zero-resistance state (ZRS) \cite{mani02,zudov03,yang03,konstantinov10} is perhaps the most spectacular manifestation of the newly discovered nonequilibrium effects in ultrahigh mobility two-dimensional (2D) electron systems in high Landau levels, driven by ac (in the microwave range) or dc electric fields \cite{dmitriev12}. ZRS is attributed \cite{andreev03} to the instability of a homogeneous state with the negative {\it absolute} dissipative conductivity $\sigma<0$ and the associated nonequilibrium phase transition into a {\it static} domain state with zero net resistance \cite{vavilov04,volkov04,dmitriev05,alicea05,finkler06,finkler09}. The domain picture is supported by a number of experiments \cite{willett04,zudov_bichrom,zudov_acdc,wiedmann10,bykov10,dorozhkin11,konstantinov12,zudov_mp}.
Similar electrical instabilities have also been known to appear in other
contexts \cite{dmitriev12,bonch-bruevich75}; most prominently, in the Gunn diode \cite{gunn63} (where most of the effects are due to the negative differential conductivity and the emergence of moving domains) and in illuminated ruby crystals \cite{ruby,dyakonov84} (where the strongly anisotropic nature of charge transport reduces the problem to one dimension, with the instability controlled by the differential conductivity). What also makes the ZRS problem special is the emergence of $\sigma<0$ in the presence of a magnetic field $B$ that produces a strong Hall component of current. While the microscopic mechanisms that lead to $\sigma<0$ in nonequilibrium 2D electron gases at $B\neq 0$ are by now fairly well established \cite{dmitriev12}, the physics of the resulting domain state remains poorly understood.

Most works \cite{andreev03,vavilov04,volkov04,dmitriev05,alicea05,finkler06,finkler09} so far have studied the bulk properties of the domain state, i.e., the limit $d/L\to 0$ in which the width $d$ of the domain wall (DW) is vanishingly small compared to the system size $L$. In fact, however, the physics related to the DW structure is crucially important near the phase transition, because the DW width diverges at the critical point and serves as a critical parameter of the transition. Moreover, the position of the phase transition in finite-size systems (nonzero $d/L$) becomes $L$ dependent and shifts towards negative $\sigma$ \cite{dorozhkin11a}.

In this paper, we develop an analytical model of the domain state for arbitrary $d/L$ and study the nonlinear response of the domain state to external voltage. The analytical solution enables us to construct the phase diagram of the biased finite-size system, which incorporates not only continuous but also discontinuous transitions between the homogeneous and domain states, and to calculate the negative conductance in the domain state.

\begin{figure}[ht]
\centerline{
\includegraphics[width=\columnwidth]{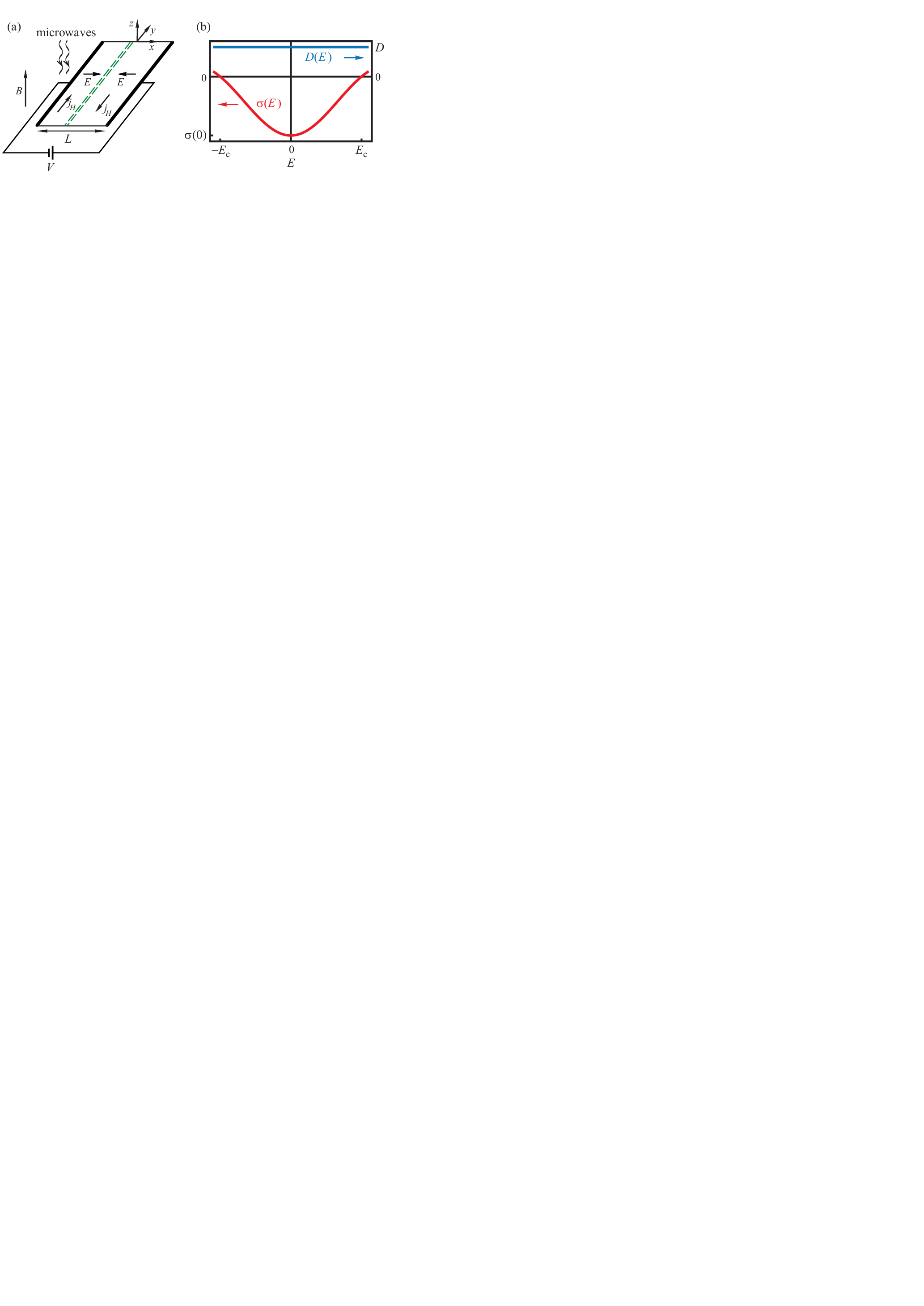}
}
\caption{(a) Geometry of the model: a 2D stripe between two metallic contacts along the long sides. The double dashed line shows a possible position of the domain wall, the arrows indicate the direction of the electric field $E$ and the Hall current $j_H$ in the voltage-biased domain state. (b) The negative nonlinear dissipative conductivity $\sigma(E)$ and the diffusion coefficient $D$ independent of $E$.
} \label{fig:model}
\end{figure}

\noindent
{\it Model.---}Consider 2D electrons occupying a stripe $(|x|< L/2,\,z=0)$, infinite in the $y$ direction (\rfig{fig:model}a), in between two plane metallic contacts at $x=\pm L/2$ that are perpendicular to the stripe. Under the illumination by microwaves at $B\neq 0$, the linear response dc dissipative conductivity $\sigma(E\to 0)$, where $E$ is the dc driving field, becomes negative in one or more intervals of $B$ for the microwave power $P$ above the critical threshold $P_c$ \cite{dmitriev12}. The nonlinear absolute conductivity $\sigma(E)$ remains negative in a finite range of $E$, crossing zero at the critical field $E_c$ (\rfig{fig:model}b) \cite{footnoteEc}. By contrast, the diffusion coefficient $D>0$ is nearly unaffected by the radiation and obeys the Einstein relation $\sigma_{\rm dark}=e^2\chi D$ valid at equilibrium, where $\sigma_{\rm dark}$ and $\chi=m/\pi$ ($m$ is the electron mass) are the dark conductivity and the dark compressibility.

We first explore the part of the phase diagram in which the system at $P\neq0$ remains homogeneous along the $y$ axis, so that the surface electron density $n_e(x)$ and the $x$ and $y$ components of the electric current $j_{x,y}(x)$ depend on $x$ only. The domain state is a stable solution of the Poisson and continuity equations. The former relates $n_e(x)$ to the normal component ${\cal E}_z(x,z)$ of the electric field at $z=\pm 0$. Namely, $2\pi e [n_e(x)-n_0]=\epsilon {\cal E}_z(x,+0)$ with $e<0$, where $n_0$ is the density of background positive charges and $\epsilon$ is the dielectric constant of the medium. The continuity equation reduces in the static limit to $\partial_x j_x(x)=0$. We assume that the in-plane component of the electric field $E(x)={\cal E}_x(x,+0)$ varies on a spatial scale  that is larger than the microscopic scales \cite{footnote_scale} of the problem and introduce the local conductivity $\sigma [E(x)]$. Importantly, at nonzero $P$, the Einstein relation does not hold and the sum of the drift and diffusion terms in the dissipative current $j_x{=}\sigma(E)E{-}e D\partial_x n_e$ is not expressible as the gradient of the electrochemical potential \cite{dmitriev09}.

At $z\neq 0$, the functions ${\cal E}_x(x,z)$ and ${\cal E}_z(x,z)$ are harmonic conjugates. This, together with the continuity of ${\cal E}_x(x,z)$ at $z=0$, allows us to represent the Poisson equation as
$\epsilon E(x)=2\pi e\,{\cal H}\{n_e(x)\}$, where the Hilbert transform ${\cal H}\{f(x)\}\equiv\pi^{-1}{\rm p.v.}\!\int dx'(x-x')^{-1}f(x')$ obeys ${\cal H}^2=-1$. Applying the Hilbert transform to the Poisson equation in this form and substituting the result in the diffusion term in $j_x$, we find that $E(x)$ satisfies the nonlinear integral equation
\be\label{main1}
E(x)\sigma [E(x)]+(\epsilon D/2\pi)\,\partial_x{\cal H}\{E(x)\}=j_x~.
\ee
Below, we solve Eq.~(\ref{main1}) with the boundary conditions $n_e(x=\pm L/2)=n_0$ and for a particular choice of
\be\label{sine-model}
\sigma(E)=\sigma(0)(E_c/\pi E)\sin(\pi E/E_c)
\ee
with $\sigma(0)<0$ (plotted in \rfig{fig:model}b) \cite{model}.

\noindent
{\it Domain solution.---}By introducing the dimensionless density $\rho(x)=\pi^2 e [n_e(x)-n_0]/\epsilon E_c$, electric field $\theta(x)=\pi E(x)/2 E_c$, and current $\tilde{j}=\pi j_x/\sigma (0) E_c$, Eqs.~(1) and (2) are rewritten in the compact form as
\be\label{main2}
\sin 2\theta+2\lambda\partial_x \rho=\tilde{j}~,
\ee
where $\rho(x)$ and $\theta(x)$ satisfy $\rho(x)=-{\cal H}\{\theta(x)\}$ and
\be\label{NSL}
\lambda=\epsilon D/2\pi|\sigma (0)|~.
\ee
Being the only spatial scale in Eq.~(\ref{main2}) at $\tilde{j}=0$, $\lambda$ is identified as the nonequilibrium screening length which gives the characteristic width of the DW \cite{footnote_scale}.

With the boundary conditions $\rho(\pm L/2)=0$, the domain solution to Eq.~(\ref{main2}) in terms of the complex function $\Psi_{\rm dom}=\theta_{\rm dom}+i\rho_{\rm dom}$ reads
\be
\Psi_{\rm dom}=i\ln\frac{\cosh(\xi-i w)}{\sinh\xi}-\frac{\arcsin\tilde{j}}{2}~,
\label{central}
\ee
where $2\xi=i\pi x/L+i w+\beta$, $w=\arctan(\tilde{j} l)$, $\beta={\rm arcoth}\sqrt{l^2-\tilde{j}^2 l^2},$ and $l=L/\pi\lambda$. The system spontaneously chooses between two degenerate domain states related to each other by $\Psi_{\rm dom}(x)\leftrightarrow\Psi_{\rm dom}^*(-x)$. For definiteness, we analyze the solution with $\theta_{\rm dom}(L/2)>0$.

\noindent
{\it Homogeneous state: Linear stability.---}For given $\tilde{j}$ in the homogeneous case of $\rho=0$, the electric field $\theta_{\rm hom}$ satisfies  $\sin 2\theta_{\rm hom}=\tilde{j}$ [Eq.~(\ref{main2})] with the solution
\bea\label{homsol}
\theta_{\rm hom}^<=(1/2)\arcsin \tilde{j}~,\qquad\theta_{\rm hom}^>=\pi/2-\theta_{\rm hom}^<~,
\eea
where the signs $\lessgtr$ correspond to the regions with the negative ($<$) and positive ($>$) differential conductivity
\be\label{sigma_d}
\sigma_d(E)=\partial_E[E\sigma(E)]=\sigma(0) \cos 2\theta~.
\ee
Linear stability analysis \cite{unpub} in the vicinity of this solution (with the electric field along the $x$ axis) shows that it is stable against small time- and space-dependent charge fluctuations proportional to $\exp(i q_xx+i q_yy)$ if
\be\label{lin_stab}
\sigma_d(E) q_x^2+\sigma(E) q_y^2>-(\epsilon D/2\pi)(q_x^2+q_y^2)^{3/2}
\ee
for all possible $q_x$ and $q_y$. In an infinite 2D system, \req{lin_stab} reduces to the usual stability conditions $\sigma_d>0$ for longitudinal ($q_y=0$, $q_x\to 0$) and $\sigma>0$ for transverse ($q_x=0$, $q_y\to 0$) fluctuations. In the stripe geometry, $q_x$ takes discrete values $\pi n/L$ with $|n|=1,2,\dots$ and the diffusion term on the right-hand side of \req{lin_stab} becomes relevant. Equation (\ref{lin_stab}) then yields
\be\label{long_stab}
\sigma_d(E) >-\epsilon D/2L
\ee
for the condition of the longitudinal stability ($q_y=0$, $|q_x|=\pi/L$). For $E\to 0$, Eq.~(\ref{long_stab}) gives the threshold value of $l=1$ \cite{dorozhkin11a} for the breakup of the homogeneous state into domains in the unbiased case, as discussed below.

\begin{figure}[t]
\centerline{
\includegraphics[width=\columnwidth]{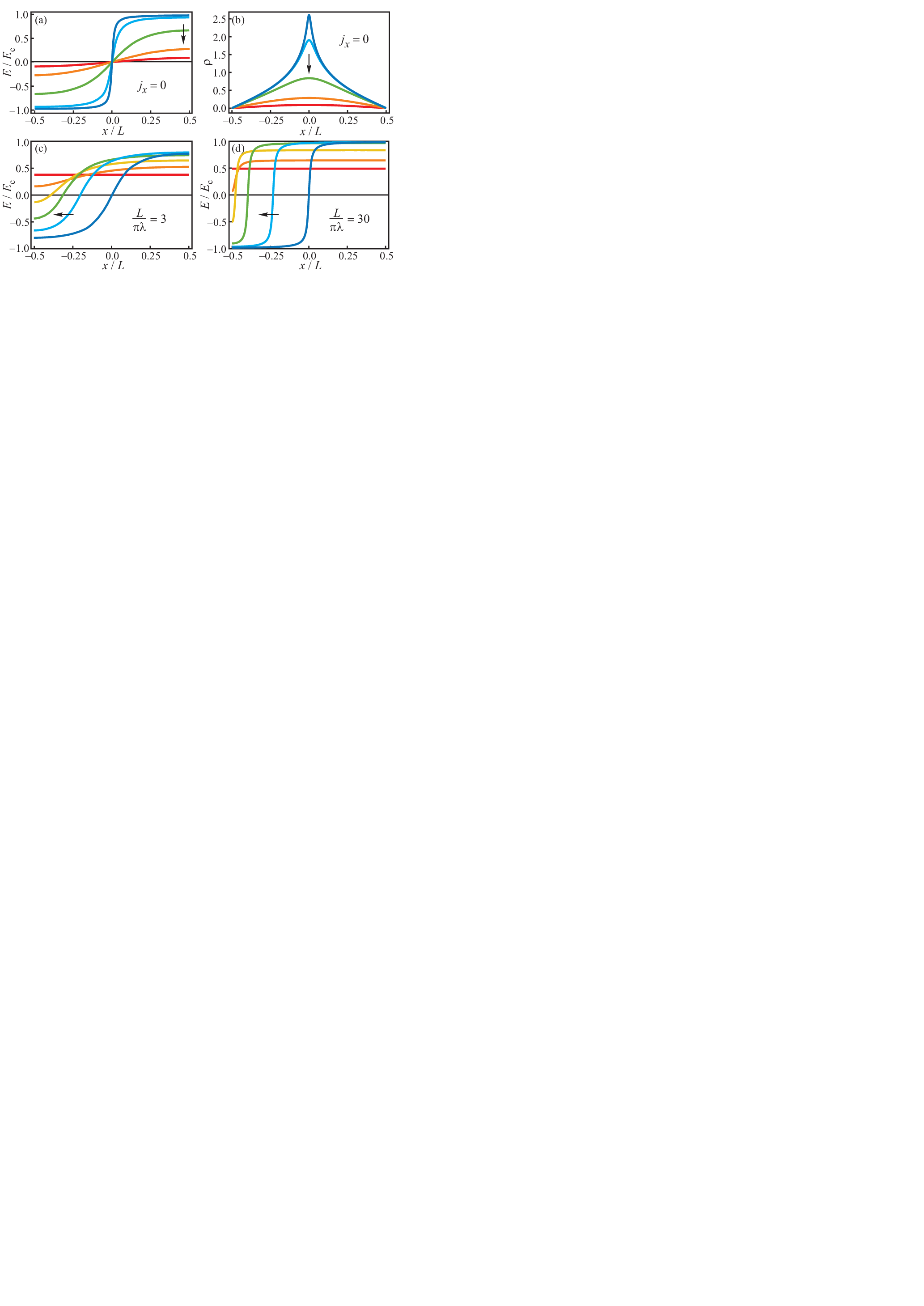}
}
\caption{Spatial distribution of (a) the electric field $E(x)$ (in units of $E_c$) and (b) the charge density $\rho(x)$ (in units of $\pi^2e/\epsilon E_c$) in the domain state of the unbiased stripe (current $j_x=0$) for $L/\pi\lambda=30,\,10,\,2,\,1.1,\,1.01$ ($L$ decreases in the direction of arrow). As the current is increased, the domain wall is shifted and broadened. The field $E(x)$ is shown for (c) $L/\pi\lambda=3$ and the current [in units of $\sigma(0)E_c/\pi$] $\tilde{j}=0,\,0.25,\,0.5,\,0.75,\,0.9,\,(8/9)^{1/2}$ and for (d) $L/\pi\lambda=30$ and $\tilde{j}=0,\,0.03,\,0.1,\,0.5,\,0.9,\,(899/900)^{1/2}$ ($\tilde j$ grows in the direction of arrow).
} \label{fig:fields}
\end{figure}

\noindent
{\it Unbiased domain state.---}For $\tilde{j}=0$,
Eq.~(\ref{central}) reduces to
\bea\label{j=0,E}
&&\theta_{\rm dom}=\arctan\,[(l^2-1)^{1/2}\sin(\pi x/L)]~,\\
\label{j=0,rho}
&&\rho_{\rm dom}={\rm artanh}\,[(1-l^{-2})^{1/2}\cos(\pi x/L)]
\eea
(\rfig{fig:fields}a,b). In the limit $l\gg 1$, Eq.~(\ref{j=0,E}) simplifies to $\theta_{\rm dom}=\arctan(x/\lambda)$, which means two domains with $E(\pm L/2)\simeq\pm E_c$ separated by the DW of width $\lambda$. The DW is charged with $\rho(x)\simeq \ln (L/|x|)$ for $\lambda\ll |x|\ll L$. This gives the oppositely directed Hall currents $j_y(x)=-en_e(x)cE(x)/B$ on the sides of the DW (Fig.~\ref{fig:model}a). With lowering $l$, both $|\theta(x)|$ and $\rho(x)$ decrease, and vanish to zero at $l=1$. As follows from Eq.~(\ref{long_stab}), the homogeneous state with $\sigma(0)<0$ in the unbiased stripe is stable for $l<1$, i.e., there is a continuous transition between the homogeneous and domain states. Near the transition, for $0< l-1\ll 1$, $\Psi_{\rm dom}\simeq i\sqrt{2(l-1)}\exp(-i\pi x/L)$ vanishes with the critical exponent $1/2$.

\noindent
{\it Biased domain state.---} For $\tilde{j}\neq 0$, Eq.~(\ref{central}) tells us that the DW shifts by $L w/\pi$ from $x=0$ while the characteristic width of the DW $d=L\beta/\pi$ grows with increasing $|\tilde{j}|$ (\rfig{fig:fields}c,d) and diverges as $(\tilde{j}_{c2}-|\tilde{j}|)^{-1/2}$ at the critical point $|\tilde{j}|=\tilde{j}_{c2}$, where
\be
\tilde{j}_{c2}=(1-l^{-2})^{1/2}~.
\label{12}
\ee
According to Eqs.~(\ref{sigma_d}) and (\ref{long_stab}), for $|\tilde{j}|>\tilde{j}_{c2}$ the homogeneous state $\theta_{\rm hom}$ is stable against longitudinal fluctuations. The line (\ref{12}) of the second-order phase transitions includes, at its endpoint at $\tilde{j}_{c2}=0$, the transition at $l=1$ in the unbiased stripe discussed above.

\begin{figure}[!]
\centerline{
\includegraphics[width=0.9\columnwidth]{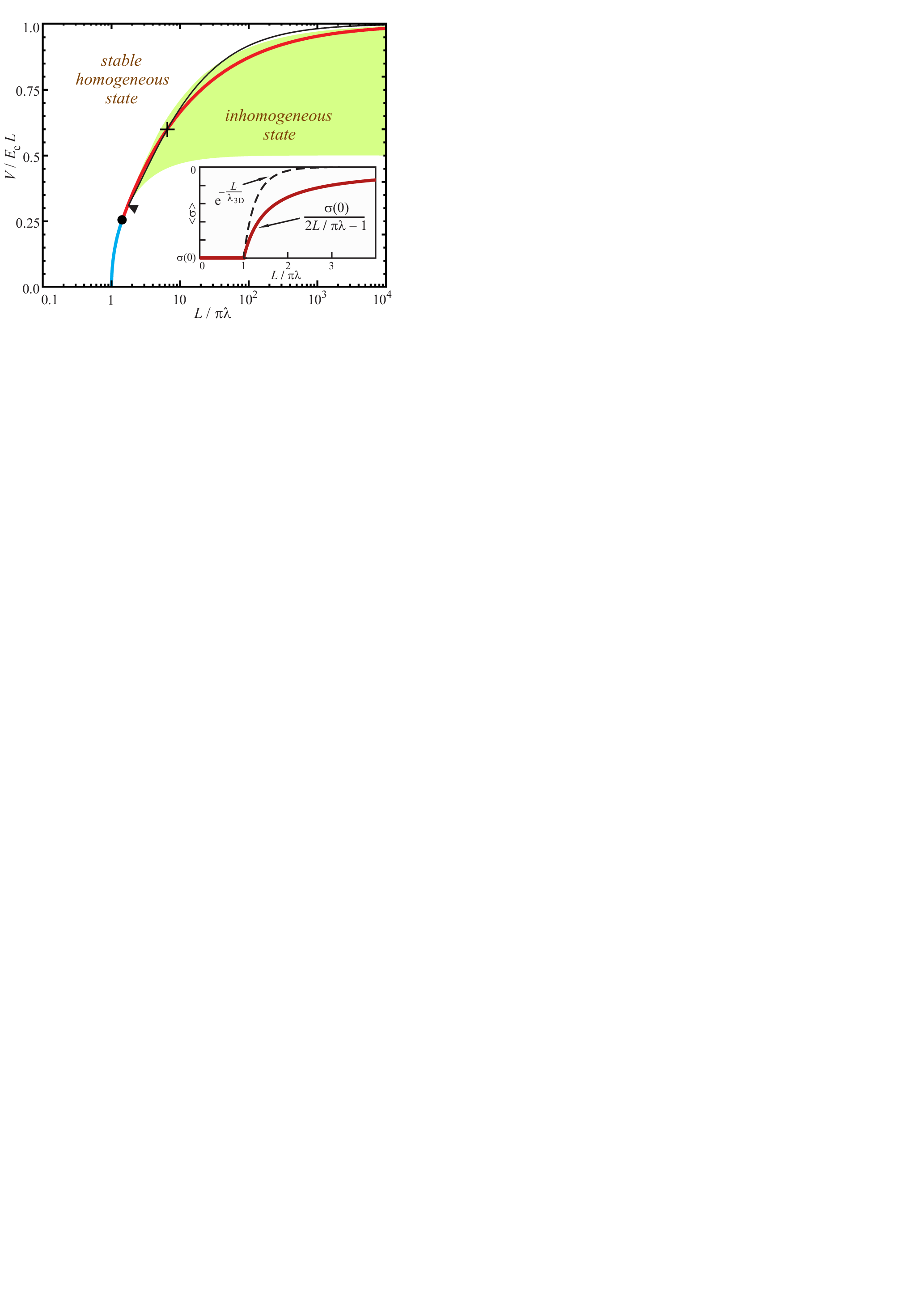}
}
\caption{Phase diagram of the voltage-biased stripe in the $V$--$L$ plane ($V$ and $L$ measured in units of $E_cL$ and $\pi\lambda$, respectively). The phase boundary (solid line), induced by the longitudinal instability, separates the homogeneous (above) and domain (below) states. In the unbiased stripe $(V\to 0$), there is a second-order phase transition at $L/\pi\lambda=1$. The continuous transition line terminates with increasing $V$ and $L$ at the tricritical point (filled circle) $V=E_cL/4$, $L/\pi\lambda=\sqrt{2}$. At larger $V$ and $L$, the transition is first order. The shaded area is the region of hysteresis. The thin line, with the endpoint marked by the triangle, denotes the linear stability threshold for the homogeneous state against transverse fluctuations (the cross marks the point at which the line intersects the discontinuous transition line obtained in the model of frozen transverse fluctuations). Inset: the effective linear-response conductivity $\langle\sigma\rangle$ as a function of $L$ [Eq.~(\ref{CVC1st})]. For comparison, the dashed line shows the behavior for a 3D model \cite{volkov04}.
}
\label{fig:phases}
\end{figure}

Averaging the field $\theta_{\rm dom}$ [Eq.~(\ref{central})] for given $\tilde{j}$ over the stripe cross-section, $\bar{\theta}_{\rm dom}=(1/L)\!\int_{-L/2}^{L/2} \!dx\,\theta_{\rm dom}(x)$, one finds the bias voltage $V=2E_cL\bar{\theta}_{\rm dom}/\pi$ and the current-voltage characteristic (CVC) of the domain state with
\be\label{CVC}
\bar{\theta}_{\rm dom}={\rm arctan}(\tilde{j} l)-(1/2)\arcsin\tilde{j}~.
\ee
Note that the current $j_x$ flows against the applied field, $j_xV<0$. For $V\to 0$, \req{CVC} gives $\bar{\theta}_{\rm dom}=(l-1/2)\tilde{j}$, or, restoring units, the linear dissipative conductance of the stripe $j_x/V=\langle\sigma\rangle/L$ in the domain state, where
\be\label{CVC1st}
\langle\sigma\rangle=\frac{\sigma(0)}{2L/\pi\lambda-1}~,\quad L>\pi\lambda~,
\ee
see inset in \rfig{fig:phases}. It is worth noting that, as $L$ increases, $\langle\sigma\rangle$ in Eq.~(\ref{CVC1st}) behaves as $L^{-1}$ in sharp contrast to the exponential behavior of $\langle\sigma\rangle\propto -\exp(-L/\lambda_{\rm 3D})$ \cite{volkov04} for a three-dimensional medium with the negative conductivity $\sigma_{\rm 3D}$, where the relation between the electric field and charge density is local. Here $\lambda_{\rm 3D}=(\epsilon D/4\pi|\sigma_{\rm 3D}|)^{1/2}$ is the analogue of $\lambda$ [\req{NSL}]. Transport across the DW is thus strongly enhanced by the nonlocal character of 2D electrostatics.

\begin{figure}[!]
\centerline{
\includegraphics[width=\columnwidth]{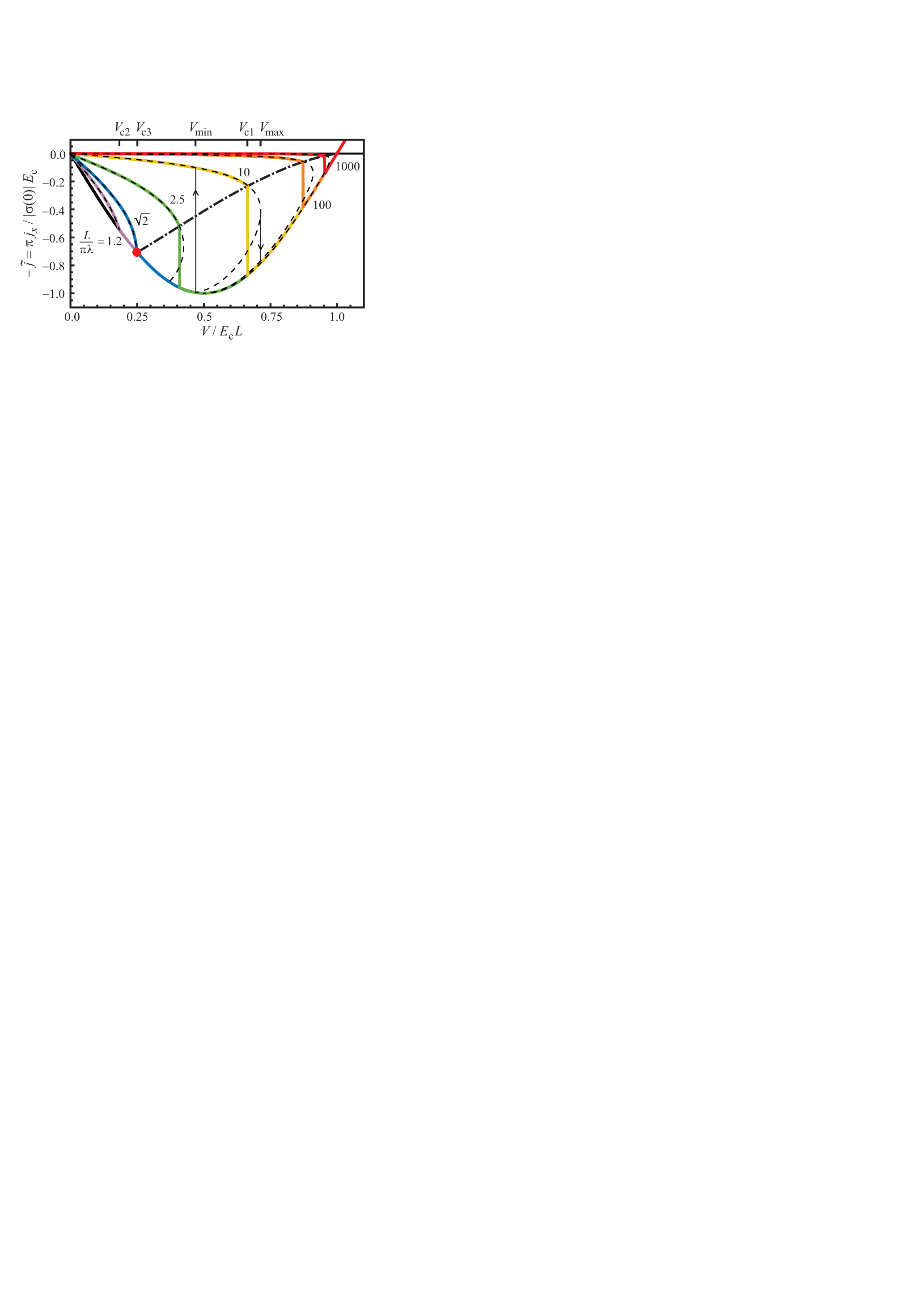}
}
\caption{Current-voltage characteristic (thick lines) of the stripe for different $L/\pi\lambda$. For $L/\pi\lambda <1$ (homogeneous state), the dependence of the dissipative current $j_x(V)$ is given by the lowest curve $\tilde{j}=\sin(\pi V/E_c L)$. For $1<L/\pi\lambda<\sqrt{2}$, the homogeneous state breaks up into domains for $V<V_{c2}$, in a continuous fashion, and the $j_x$--$V$ curve has a kink, as shown for $L/\pi\lambda=1.2$. For $L/\pi\lambda>\sqrt{2}$, the $j_x$--$V$ curves for the domain state (dashed lines) are double-valued and the transition becomes discontinuous, as demonstrated by the jumps of $j_x(V)$ at $V=V_{c1}$ (the critical voltage is marked for $L/\pi\lambda=10$) in the curves for $L/\pi\lambda=2.5,\,10,\,100,\,1000$. The discontinuous transition line (dash-dotted) terminates at the tricritical point (filled circle) at $V=V_{c3}$. The arrows on the thin vertical lines denote hysteresis (shown for $L/\pi\lambda=10$) in the interval $V_{\rm min}<V<V_{\rm max}$.} \label{fig:cvc}
\end{figure}

The CVC (\ref{CVC}) for several values of $l$ is shown by dashed lines in Fig.~\ref{fig:cvc}. For $l>\sqrt{2}$, the current is seen to become, as the voltage is increased, a double-valued function of $V$. That is, in fact, the continuous transition line (\ref{12}) in the $V$--$l$ plane terminates at $l=\sqrt{2}$ and becomes first order for larger $l$ (\rfig{fig:phases}), as we discuss next.

\noindent
{\it Lyapunov functional.---}The linear stability analysis, which leads to Eq.~(\ref{long_stab}) and the continuous transitions in the interval $1<l<\sqrt{2}$, does not capture the emergence of the discontinuous transitions for $l>\sqrt{2}$, i.e., [by substituting $l=\sqrt{2}$ in Eqs.~(\ref{12}) and (\ref{CVC})] for large voltages $V>E_cL/4$. The stability analysis of the domain solution (\ref{central}) that we perform below to describe the large-voltage regime is based on the Lyapunov functional (LF) approach to the ZRS problem \cite{finkler06}. The advantage of the LF method is that it is capable of discriminating the stable (global minimum of the LF) and metastable (local minimum) states that can be distant in phase space. The LF $\Phi\{E(x)\}= -G + K$ is given by the difference of the gain $G=-\int dx \!\int^{E(x)}_0\! dE' E' \sigma(E')$ and the DW contribution $K= (D/2) \int \!dx\, E(x) \hat{C} E(x)$, where the capacitance operator $\hat{C}=(\epsilon/2\pi)\partial_x \cal{H}$. For the model (\ref{sine-model}), we have
\be\label{lyap_sin}
\Phi\!=\!-\!\int\!\frac{dx}{L}\sin^2\!\theta(x)+\!\lambda\int\! \frac{dx\,dx'}{2\pi L}\,\left[\frac{\theta(x)\!-\!\theta(x')}{x-x'}\right]^2~,
\ee
which gives $\Phi_{\rm hom}=-\sin^2\theta_{\rm hom}$ in the uniform state and
\be\label{Phi_dom}
\Phi_{\rm dom}=-\Big(1+\sqrt{1-\tilde{j}^2}\Big)/2+l^{-1}-l^{-1}\ln\sqrt{\tilde{j}^2+l^{-2}}
\ee
in the domain state (\ref{central}) \cite{absmin}.

\noindent
{\it First-order transitions.---}Comparison of $\Phi_{\rm hom}$ and $\Phi_{\rm dom}$ leads to the phase boundary (thick line) in Fig.~\ref{fig:phases} and the CVC (thick lines) for different $l$ in Fig.~\ref{fig:cvc}. For $l<1$, the homogeneous state is stable ($\Phi_{\rm hom}<\Phi_{\rm dom}$) for arbitrary $V$. For a given $l$ in the interval $1<l<\sqrt{2}$, there is a continuous voltage-driven transition ($\Phi_{\rm hom}=\Phi_{\rm dom}$) between the homogeneous and domain states at $V=V_{c2}\equiv (E_c L/\pi)\arccos l^{-1}$. For $l>\sqrt{2}$, there emerges the interval $V_{\rm min}<V<V_{\rm max}$ (whose endpoints are shown in \rfig{fig:cvc} for $l=10$) in which the function $\tilde{j}(V)$ for the domain state is double-valued (dashed lines in \rfig{fig:cvc}). Note that the expression for $V_{\rm min}$ coincides with that for $V_{c2}$, so that $V_{\rm min}$ saturates as $l$ is increased at $E_cL/2$. On the lower branch (with $dV/d\tilde{j}<0$), $\Phi_{\rm dom}$ is larger than on the upper (and the lower branch is unstable against linear fluctuations \cite{unpub}). For the upper branch, the phase boundary equation $\Phi_{\rm hom}=\Phi_{\rm dom}$ (whose solution is shown in Fig.~\ref{fig:cvc} as a dash-dotted line) yields, for $l>\sqrt{2}$, the first-order transition at $V=V_{c1}$, where $V_{c1}$ tends to $E_cL$ in the ``bulk" limit $l\to\infty$ as $V_{c1}/E_cL\simeq 1-(1/\pi)(2\ln l/l)^{1/2}$ \cite{1order}. The current discontinuity [which vanishes for $\tilde j$ as $(2\ln l/l)^{1/2}$ for $l\gg 1$], between the upper-branch domain state and the homogeneous state, is illustrated by the vertical thick lines in Fig.~\ref{fig:cvc}. The ``tricritical" point, separating the first- and second-order transitions, at $l=\sqrt{2}$ and $V=V_{c3}\equiv E_cL/4$ is marked by the filled circle in Figs.~\ref{fig:phases} and \ref{fig:cvc}.

The domain state for $V_{c1}<V<V_{\rm max}$ and the homogeneous state for $V_{\rm min}<V<V_{c1}$ are metastable, i.e., they can be probed if the voltage sweep rate is larger than their characteristic decay rates (the dynamical properties of the system are beyond the scope here: in the above, we assumed that the system resides in the stable state, in which $\Phi$ is minimized globally). In the nonadiabatic limit, the system exhibits hysteresis \cite{dya}, as marked in Fig.~\ref{fig:cvc} by the arrows on the thin vertical lines for the case of $l=10$. The hysteresis range $V_{\rm min}<V<V_{\rm max}$ for arbitrary $l>\sqrt{2}$ is shown as a shaded area in Fig.~\ref{fig:phases}.

\noindent
{\it Transverse instability.---}Before concluding, we briefly comment on the stability of the above picture against transverse fluctuations. Allowing for nonzero $q_y$ in \req{lin_stab}, they can be shown \cite{unpub} to be irrelevant on the phase boundary (thick line) in Fig.~\ref{fig:phases} for a sufficiently narrow stripe, namely for $l<l_{c\perp}\simeq 1.76$ (or, equivalently, $V<V_{c\perp}\simeq 0.31E_cL$). The threshold is marked in Fig.~\ref{fig:phases} by the triangle. The range of $l<l_{c\perp}$ includes the zero-bias critical point at $l=1$ and the tricritical point at $l=\sqrt{2}$. For $l>l_{c\perp}$, the linear transverse stability is maintained above the thin line in Fig.~\ref{fig:phases}, which runs well above the longitudinal stability threshold $V_{\rm min}$
for the homogeneous state (\ref{long_stab}) and very closely to the discontinuous transition line $V_{c1}$ obtained for frozen transverse fluctuations. At $l=l_{c+}\simeq 7$ (marked by the cross in Fig.~\ref{fig:phases}), the two lines intersect, so that for $l>l_{c+}$ the homogeneous state is unstable above the phase boundary at $V_{c1}$. In the narrow region between the lines for $l>l_{c+}$, the global minimum of the LF should thus be given by a 2D domain state with broken translational invariance along the stripe. The nature of this state, as well as the position of the boundary $V_{c1}^*$ for the global stability of the domain state (\ref{central}) \cite{lin_stab}, requires additional study.

\noindent
{\it Summary.}---We have studied transport in the voltage-biased stripe with a negative absolute conductivity and obtained the phase diagram which shows phase transitions between the domain and homogeneous states. The transitions are second order for small and first order for large voltages (Fig.~\ref{fig:phases}). We have calculated the current-voltage characteristic of the domain state (Fig.~\ref{fig:cvc}) and found the negative dissipative conductance. Our predictions can be verified by measuring the current-voltage characteristic in sufficiently small samples 
in the vicinity of the ZRS transition.

\noindent
{\it Acknowledgements.}---We thank S.~Dorozhkin, Y.~Galperin, A.~Kamenev, P.~Ostrovsky, I.~Protopopov, J.~Smet, R.~Suris, and M.~Zudov for interesting discussions.
The work was supported by the DFG-RFBR (I.A.D. and A.D.M.) and by the University of Iowa (M.K.).


\begin{thebibliography}{99}
\bibitem{footnote_scale} Using the local conductivity is justified if the nonequilibrium screening length $\lambda$ [Eq.~(\ref{NSL})] is larger than the microscopic scales (the cyclotron radius $R_c$ and the inelastic scattering length $l_{\rm in}$ \cite{dmitriev12}). Note that $\lambda$ diverges at $\sigma(0)\to 0$, thus serving as the critical parameter for the transition to ZRS in the bulk limit $L/\lambda\to\infty$. Deep in the domain state, for $|\sigma(0)|\sim\sigma_{\rm dark}$, the screening length is of the order of $\lambda_{\rm dark}=\lambda\vert_{\sigma(0)=\sigma_{\rm dark}}=\epsilon/2me^2$. In typical experimental conditions, $\lambda_{\rm dark}\ll l_{\rm in},R_c$ and the width of the domain wall is determined by $l_{\rm in}$ and $R_c$.
\bibitem{model} This model function allows for an exact analytical solution to Eq.~(1), while the physics the solution reveals is insensitive to the exact shape of the single-parameter function $\sigma(E)$. Note that the behavior of $\sigma(E)$ for $|E|>E_c$ plays no role in what follows.
\bibitem{dya} Note that the obtained hysteresis bears close similarity to the nonequilibrium first-order phase transition proposed in Ref.~\cite{dyakonov84} to describe the experiment \cite{ruby} on illuminated ruby crystals. The key differences between our problem and that in Ref.~\cite{dyakonov84} are the importance of transverse fluctuations, which are absent in the model of Ref.~\cite{dyakonov84}, and the nonlocal character of 2D electrostatics. We also found a richer behavior of the system by studying the case of smaller system sizes, $L/\pi\lambda \sim 1$, where there emerge the continuous phase transitions and the tricritical point that separates the regimes of the first- and second-order phase transitions.
\bibitem{absmin} Note that the absolute minimum $\Phi=-1$ of the LF (\ref{lyap_sin}) is realized in the state $|\theta(x)|=\pi/2$, i.e., in a system of size $L$,  only at $|V|=E_c L$.
\bibitem{1order} It is worth emphasizing that our consideration was performed in the mean-field framework. The impact of fluctuations on this problem remains to be studied. In particular, one can apply the methods of Ref.~\cite{alicea05} where it was found (in the limit $V\to 0$ and $L\to\infty$) that fluctuations may force the ZRS transition to be first order. Remarkably,  we obtain the first-order transition driven by finite voltage already on the mean-field level.
\bibitem{lin_stab} We were able to check numerically that the linear stability threshold of the domain state (\ref{central}) remains above $V_{c1}$, i.e., the state (\ref{central}) is linearly stable in the whole region $V<V_{c1}$, at least for $l<40$.
\end{thebibliography}
\end{document}